\documentclass[alpha-refs]{wiley-article}

\usepackage{color}
\usepackage{ulem}

\papertype{Original Article}
\paperfield{Journal Section}

\title{The impact of decreasing horizontal grid spacing on the simulation of the mountain boundary layer in the hectometric range}

\author[1\authfn{1}]{Brigitta Goger}
\author[1\authfn{1}]{Anurag Dipankar}

\affil[1]{Center for Climate Systems Modeling (C2SM), ETH Zurich, Zurich, Switzerland}

\corraddress{Brigitta Goger, Center for Climate Systems Modeling, ETH Zurich, Zurich, Switzerland}
\corremail{brigitta.goger@c2sm.ethz.ch}

\fundinginfo{ETH Zurich}

\runningauthor{Goger and Dipankar}

\begin{document}

\maketitle

\begin{abstract}
The horizontal grid spacing of numerical weather prediction models keeps decreasing towards the hectometric range. We perform limited-area simulations with the ICON model across horizontal grid spacings (1\,km, 500\,m, 250\,m, 125\,m) in the Inn Valley, Austria, and evaluate the model with observations from the CROSSINN measurement campaign. This allows us to investigate whether increasing the horizontal resolution automatically improves the representation of the flow structure, surface exchange, and common meteorological variables. Increasing the horizontal resolution results in an improved simulation of the thermally-induced circulation. However, the model still faces challenges with scale interactions and the evening transition of the up-valley flow. Differences between two turbulence schemes (1D TKE and 3D Smagorinsky) emerge due to their different surface transfer formulations, yielding a delayed evening transition in the 3D Smagorinsky scheme. Generally speaking, the correct simulation of the mountain boundary layer depends mostly on the representation of model topography and surface exchange, and the choice of turbulence parameterization is secondary.
\end{abstract}

\section{Introduction}
Mountainous terrain strongly influences the exchange of heat, mass, and momentum between the surface and the atmosphere \citep{RotachEtAl_2022_CollaborativeEffortBetter}. The underlying surface heterogeneity leads to the formation of the so-called mountain boundary layer \citep{LehnerRotach_2018_CurrentChallengesUnderstanding}, characterized by complex flow structures on several length and time scales interacting with each other. For example, larger-scale dynamically-induced foehn flow removes a cold-air pool in a valley \citep{HaidEtAl_2020_Foehncoldpool}, the up-valley flow erodes smaller-scale slope flows \citep{RotachEtAl_2008_Boundarylayercharacteristics}, or the sea breeze interacts with local valley circulations \citep{BantaEtAl_2023_MeasurementsModelImprovement}. \par
With the rise of computational power in recent years, operational numerical weather prediction (NWP) models operate at the kilometric range \citep{BauerEtAl_2015_quietrevolutionnumerical,SchmidliEtAl_2018_AccuracySimulatedDiurnal}. Recent efforts by operational centres aim for a jump towards hectometric grid spacings ($\Delta x=\mathcal{O}{[100\,m]}$), e.g., the GLORI Digital Alpine Twin Project \citep{glori} and current developments by M\'{e}t\'{e}o France \citep{Honnert_2022_HectometricScalesinNWP}. Higher resolution also leads to a more realistic representation of model topography, and simulations over idealized topography revealed that at least ten grid points across a valley are necessary to resolve mountain boundary-layer processes accordingly in models \citep{WagnerEtAl_2014_ImpactHorizontalModel}. This criterion is met by NWP models at the kilometric range already for major Alpine valleys (e.g., the Inn Valley in Austria or the Swiss Rhone Valley) with a successful simulation of daytime up-valley flows \citep{SchmidliEtAl_2018_AccuracySimulatedDiurnal,MikkolaEtAl_2023_Daytimevalleywinds}. However, real mountainous terrain also consists of smaller tributary valleys and basins usually not resolved at $\Delta x=1$\,km. Therefore, a decrease of $\Delta x$ towards the hectometric range is likely not only beneficial for the simulation of clouds and precipitation as suggested by \cite{StevensEtAl_2020_AddedValueLarge}, but also for an improved representation of atmospheric processes over truly complex terrain. \par
However, due to the complex character of the mountain boundary layer \citep{LehnerRotach_2018_CurrentChallengesUnderstanding}, classical boundary-layer theory assumptions (e.g., horizontally homogeneous and flat terrain) used in physical paramterizations are violated \citep{ZhongChow_2013_MesoFineScale,MunozEsparzaEtAl_2014_BridgingTransitionMesoscale,RotachEtAl_2017_InvestigatingExchangeProcesses,StiperskiCalaf_2023_GeneralizingMoninObukhov}. This makes a transition towards the hectometric range challenging, because physical parameterizations have to be adjusted accordingly. For example, the turbulence scheme has to transition from 1D parameterizations \citep[e.g., classical schemes like][]{MellorYamada_1982_Developmentturbulenceclosure} towards more complex schemes, e.g., blending a 1D parameterization and a large-eddy simulation (LES) closure \citep{ZhangEtAl_2018_ThreeDimensionalScale,Efstathiou_2023_DynamicSubgridTurbulence}, to tackle the challenges of the turbulence grey zone, where turbulence is party parameterized and resolved \citep{Wyngaard_2004_NumericalModelingTerra, HonnertEtAl_2020_AtmosphericBoundaryLayer}. In complex terrain, including three-dimensional effects such as horizontal shear leads to an improved simulation of the turbulence structure \citep{ZhongChow_2013_MesoFineScale,GogerEtAl_2018_ImpactThreeDimensional,GogerEtAl_2019_NewHorizontalLength,JulianoEtAl_2022_GrayZoneSimulations}. However, the choice to switch to fully three-dimensional turbulence treatments depends on length scale of the phenomena of interest and boundary layer type \citep{HonnertMasson_2014_Whatissmallest,Cuxart_2015_WhenCanHigh}. \par
\cite{ZhouEtAl_2014_ConvectiveBoundaryLayer} stress that despite the challenges arising due to the turbulence grey zone, simulations over complex terrain at the hectometric range are worth pursuing due to the higher resolution of topography and land-use, leading to a better representation of surface exchange. Furthermore, for high-resolution simulations, the correct representation of surface properties of soil and land-use with equally high-resolution datasets is crucial for a realistic boundary layer representation over mountains \citep{KalverlaEtAl_2016_EvaluationWeatherResearch,GolzioEtAl_2021_LandUseImprovements,GogerEtAl_2022_Largeeddysimulations}. The correct evolution of the local flow structure depends strongly on the correct representation of the Bowen ratio \citep{RihaniEtAl_2015_Isolatingeffectsterrain,ChowEtAl_2006_HighResolutionLarge}. \cite{SchmidliQuimbayoDuarte_2023_DiurnalValleyWinds} showed in a recent study in the Swiss Rhone Valley that a decrease of $\Delta x$ to 550\,m results in a better simulation of the up-valley flows compared to the kilometric range - but only in combination with improved soil moisture initialization.  Furthermore, \cite{TianEtAl_2024_stationbasedevaluation} found that switching from kilometric simulations to $\Delta x$=550\,m leads to a reduction of a cold bias connected with dynamically-induced foehn flows. 
\par
Another model challenge to consider is the validation of NWP models at the hectometric range, because standard model validation methods have to be improved or adjusted with increasing model resolution \citep{DorningerEtAl_2018_SetupMesoVICTProject}. The classical grid point-by-grid point model validation is likely unsuitable for mountainous terrain, because the question on representativeness of a single station in time and space arises. Furthermore, single point observations cannot asses whether the numerical model is able to account for heterogeneity in space. This problem can be partly overcome by arranging arrays of measurement stations to assess spatial variability of phenomena \citep{RotachEtAl_2017_InvestigatingExchangeProcesses,GogerEtAl_2018_ImpactThreeDimensional,GerberEtAl_2018_Spatialvariabilitysnow,RohanizadeganEtAl_2023_HighResolutionLarge}. Still, most detailed information on the boundary layer can be obtained via measurement campaigns \citep{BantaEtAl_2013_ObservationalTechniquesSampling} with extensive, multi-dimensional measurements networks consisting of turbulence flux towers, LIDARs, temperature profilers, scintillometers, etc. LIDAR observations are especially useful, because their scanning strategies (e.g., vertical profiles or co-planar scans) provide two-dimensional information on the ABL flow structure. Therefore, given the detailed observed flow structures, the meteorological fields from high-resolution models can be validated in detail \citep{HaldEtAl_2019_LargeEddySimulations,BauerEtAl_2023_EvolutionConvectiveBoundary}; e.g., the evolution and build-up of a sea-to-mountain breeze circulation in the Western US in different versions of the HRRR model was validated with multiple LIDAR systems \citep{BantaEtAl_2023_MeasurementsModelImprovement}. Furthermore, observational data from measurement campaigns also allow a process-based model validation, i.e., allowing to identify the physical processes which cause possible model deficiencies.
\par
Several real-case studies of simulations at the hectometric range over complex terrain exist, mostly for selected events \cite[e.g.,][]{ChowEtAl_2006_HighResolutionLarge,HeinzeEtAl_2017_Largeeddysimulations}.  \cite{BauerEtAl_2023_EvolutionConvectiveBoundary} performed simulations at the hectometric range with the WRF model over flat terrain and found mostly an improved representation of vertical velocity and 2\,m temperature. Nested set-ups over complex terrain exist, and suggest that higher-resolution runs lead to a better simulation of, e.g., horizontal wind speed or 2\,m temperature \citep{GerberEtAl_2018_Spatialvariabilitysnow,LiuEtAl_2020_SimulationFlowFields}, but the focus of these studies was on other research questions. To our current knowledge, no study of a \textit{systematic} model evaluation across several grid spacings in the hectometric range exists with a focus on flow structures  over complex topography. \par
 In this study, we employ a nested set-up of the Icosahedral Nonhydrostatic (ICON) Weather and Climate Model across several grid spacings across the hectometric range (1\,km--125\,m) over a truly complex terrain. The area of interest is the Inn Valley, Austria, located in the Alps, where the CROSSINN measurement campaign took place in summer and autumn of 2019 \citep{AdlerEtAl_2021_CROSSINNFieldExperiment}. We test the model for the standard NWP configuration that uses TKE based one-dimensional boundary layer scheme (1D TKE) against the research configuration that uses the three-dimensional (3D) Smagorinsky-type scheme. The 3D Smagorinsky is chosen because it helps contrast the added value of including three-dimensionality in turbulence representation, as initially intended by \cite{Smagorinsky_1963_Generalcirculationexperiments}. We aim to report on the current performance of a state-of-the-art NWP model in the hectometric range over complex topography, and try to understand (i) if higher resolution automatically improves the simulation of the thermally-induced flow structure in the valley, (ii) which physical processes are responsible for possible model deficiencies, and (iii) whether the chosen turbulence scheme configuration has an impact on the simulated boundary-layer flow structure.

\section{Data and Methods}\label{sec2}
\subsection{Observations}
\begin{figure}
 \centering
\noindent\includegraphics[width=.75\textwidth,keepaspectratio]{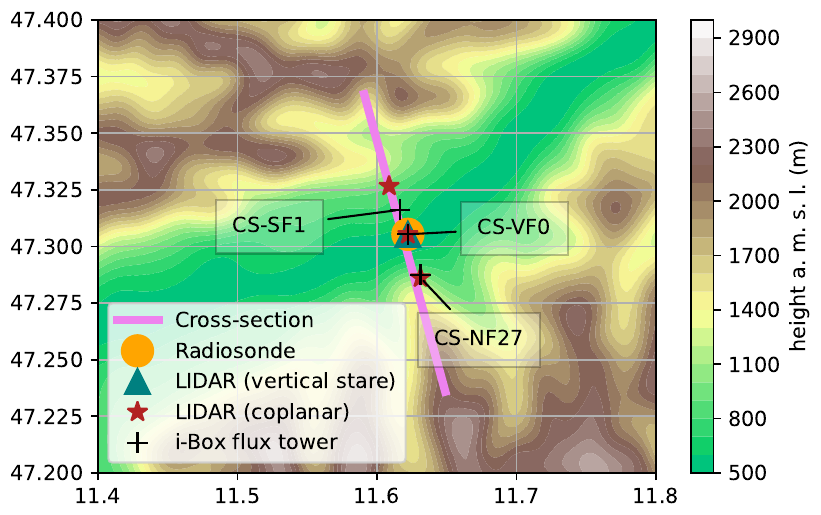}
\caption{Overview of the measurement sites used in this study. The contour colors show the terrain, while the color dots represent the kind of observation (LIDAR, radiosonde, i-Box station).} 
\label{f:stations}
\end{figure}

Our area of interest is located in the Inn Valley in the Austrian Alps, a large
East-West oriented valley, with a peak-to-peak distance of around 10\,km at our area of interest (Fig~\ref{f:stations}). The valley is located in the Central European Time Zone (UTC+1). The Inn Valley was subject to a rich body of mountain boundary layer research and will be a main target area for the upcoming TEAMx measurement campaign \citep{RotachEtAl_2022_CollaborativeEffortBetter}. The Inn Valley is a primary location for strong up-valley flows \citep{VergeinerDreiseitl_1987_Valleywindsslope,LehnerEtAl_2019_MethodIdentifySynoptically} and therefore an excellent location to investigate this mountain boundary layer phenomenon. The ``i-Box" turbulence flux towers \citep{RotachEtAl_2017_InvestigatingExchangeProcesses}, located 30\,km East of the city of Innsbruck, are operational since 2013 and allow a process-based model validation beyond standard model verification methods \citep{GogerEtAl_2018_ImpactThreeDimensional,GogerEtAl_2019_NewHorizontalLength}.  \par
In summer and autumn of 2019, the Cross-Valley Flow in the Inn Valley Investigated by Dual-Doppler LIDAR Measurements (CROSSINN) campaign took place at the location of the i-Box stations \citep{AdlerEtAl_2021_CROSSINNFieldExperiment}, investigating besides the along-valley flows also mechanisms behind cross-valley circulations \citep{BabicEtAl_2021_Crossvalleyvortices}. The operational i-Box eddy covariance flux towers were accompanied by remote sensing systems such as LIDARs in vertical stare mode and coplanar scan modes, a HATPRO temperature profiler, radiosondes, and aircraft observations during intensive observation periods (IOPs). 
The location of the measurement sites is shown in Figure~\ref{f:stations}: We will mostly utilize the collected data from the LIDAR observations, namely from the vertically pointing LIDARS SLXR 142 and SL 88 at the valley floor site (CS-VF0) and the coplanar scans across the valley \cite[their Figure 1]{BabicEtAl_2021_Crossvalleyvortices} retrieved from LIDARs located at CS-VF0, CS-SF11, and CS-NF27. Furthermore, use observations from three i-Box stations along the valley cross-section, namely from the valley floor (CS-VF0), the North-facing slope (CS-NF27), and the South-facing Slope (CS-SF1).  \par

\subsection{Numerical Model}
 \begin{figure}
 \centering
 \noindent\includegraphics[width=.75\textwidth,keepaspectratio]{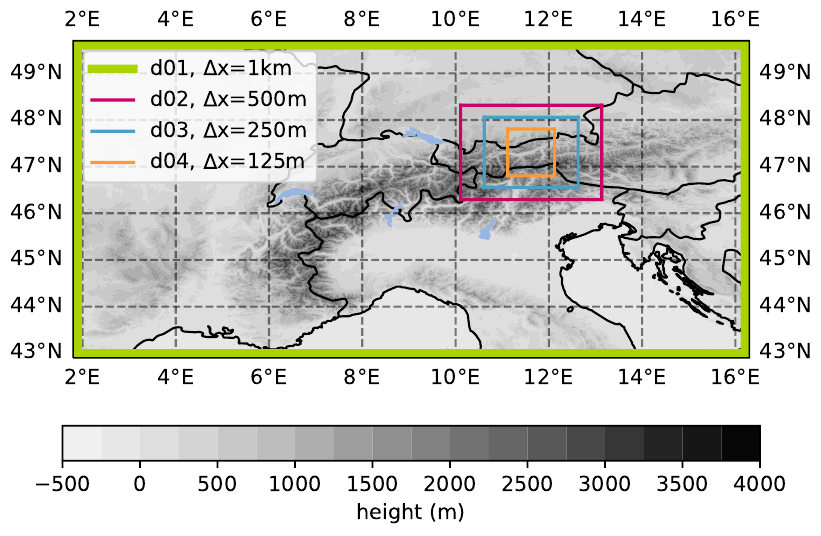}
\caption{Grey contours: Model topography from the outermost domain at $\Delta x$=1\,km. The boxes in color represent the location of the four model domains centered at the Inn Valley, Austria.}
\label{f:domains}
\end{figure}
We employ a nested set-up of the ICOsahedral Non-hydrostatic (ICON) Weather and Climate Model version 2.6.5 \citep{ZaenglEtAl_2015_ICONICOsahedralNon}. As boundary conditions, we utilize ECMWF's IFS forecast fields every hour, and we use MeteoSwiss' COSMO-1 analysis fields ($\Delta x=1$\,km) as initial conditions \citep{SchmidliEtAl_2018_AccuracySimulatedDiurnal}. The model set-up contains four nested domains (Fig.~\ref{f:domains}, Tab.~\ref{t:sim_overview}) with horizontal grid spacings of $\Delta x=1$\,km (DX1000), $\Delta x=500$\,m (DX500), $\Delta x=250$\,m (DX250), and $\Delta x=125$\,m (DX125). The grid refinement procedure at the domain boundaries in ICON follows \cite{ZaenglEtAl_2022_GridrefinementICON}. The outermost domain (DX1000) spans the Alps to ensure a correct representation of the synoptic situation over the Alps, while we subsequently nest down the domains by a factor two towards the domain of DX125, centered over the Eastern part of the Inn Valley, Austria (Fig.~\ref{f:domains}). All domains use the same 80 vertical levels in terrain following SLEVE coordinates \citep{LeuenbergerEtAl_2010_GeneralizationSLEVEVertical}, with the lowest model level being located 20\,m above ground. All simulations are started at 00:00\,UTC of the respective case study day and are run for 24\,hours in one-way nesting mode, while we regard the first three hours of simulation as model spin-up as in \cite{GogerEtAl_2022_Largeeddysimulations}. The dates for the case studies are 04 Aug 2019, 14 Aug 2019, 30 Aug 2019, 13 Sept 2019, and 14 Sept 2019. \par
 \begin{table}
 \caption{Overview of simulations conducted in this study.}
 \label{t:sim_overview}
 \centering
 \begin{tabular}{l | c | c}
 \hline
   Turbulence scheme & Simulations & horizontal grid spacing (m) \\
 \hline
   1D TKE \citep{Raschendorfer_2001_newturbulenceparameterization}  & DX1000-1D &  1000 \\
 & DX500-1D &  500 \\
  & DX250-1D &  250 \\
   & DX500-1D &  125 \\
   \hline
   3D Smagorinsky \citep{DipankarEtAl_2015_Largeeddysimulation}  & DX1000-3D &  1000   \\
 & DX500-3D &  500 \\
  & DX250-3D &  250 \\
   & DX500-3D &  125 \\
 \hline
 \end{tabular}
\end{table}
Since the model grid spacing is located at the hectometric range, high-resolution static input data is necessary to ensure realistic boundary layer development. For topography, we employ the ASTER dataset with a horizontal resolution of 30\,m \citep{NSUAST_2009_ASTERGlobalDigital}. 
ICON can run without numerical instabilities with slope angles steeper than 40$^{\circ}$ \citep{Zaengl_2012_ExtendingNumericalStability}, therefore no topographic filtering was necessary for the DX1000 and DX500 domains in our simulations, but ICON's built-in filtering algorithm (a $\nabla^2$ diffusion operator) was applied for the high-resolution domains DX250 and DX125. Still, maximum slope angles in all domains range over 45$^{\circ}$, ensuring realistic representation of topograhy in the model. \par 
For land use representation, most ICON model set-ups use the GLOBCOVER2009 dataset \citep{HeinzeEtAl_2017_Largeeddysimulations,SinghEtAl_2021_Sensitivityconvectiveprecipitationa}. We chose the CORINE land use dataset instead \citep{EEA_2017_CopernicusLandService}, because test simulations with CORINE yielded improved model performance for the thermally-induced flows in the Inn Valley over GLOBCOVER. The soil is represented by the Harmonized World Soil database  \citep{FAO/IIASA/ISSCAS/JRC_2012_HarmonizedWorldSoil}. ICON is coupled to the multi-layer land surface scheme \texttt{TERRA\_ML} \citep{Schulz_terra} and the lake model \texttt{flake} \citep{Mironov_flake}. 
We utilize the ECRad scheme \citep{HoganBozzo_2016_ECRADnewradiation} as the radiation parameterization and a single-moment microphysics scheme after \cite{Seifert_2006_RevisedCloudMicrophysical} including cloud ice, snow, and graupel. Since our ICON set-up operates at horizontal grid spacings below 1\,km, both the deep and shallow convection schemes are switched off to maintain physical consistency across resolutions. \par
We conduct two sensitivity runs for all case study days with two different turbulence parameterizations available in ICON (Tab.~\ref{t:sim_overview}):
\begin{enumerate}
    \item \textbf{1D TKE} The standard turbulence scheme in ICON is \texttt{turbdiff} developed by \cite{Raschendorfer_2001_newturbulenceparameterization} based on \cite{MellorYamada_1974_HierarchyTurbulenceClosure,MellorYamada_1982_Developmentturbulenceclosure}. The scheme considers the vertical turbulent exchange with turbulence kinetic energy (TKE) as a prognostic variable \citep{GoeckeMachulskaya_2021_AviationTurbulenceForecasting}. The scheme is used operationally at the German Weather Service on global and regional scales \citep{ZaenglEtAl_2015_ICONICOsahedralNon}. 
    \item \textbf{3D Smagorinsky} A Smagorinsky-type closure was implemented in ICON by \cite{DipankarEtAl_2015_Largeeddysimulation} following \cite{Smagorinsky_1963_Generalcirculationexperiments} and \cite{LILLY_1962_numericalsimulationbuoyant}. This scheme assumes fully isotropic 3D turbulence, while the turbulence length scale depends on the Smagorinsky constant and the grid spacing $\Delta x$. A large-scale evaluation study of the 3D Smagorinsky scheme at $\Delta x=156$\,m over Germany by \cite{HeinzeEtAl_2017_Largeeddysimulations,SchemannEtAl_2020_LinkingLargeEddy} revealed that ICON is able to simulate realistic boundary layer development at the hectometric range. 
    
\end{enumerate}

It is to be noted that the two boundary layer schemes use different schemes for surface exchange. 1D TKE is coupled to a scheme, as outlined in \cite[Appendix B]{Buzzi_2008_ChallengesOperationalWintertime}, that uses the prognostic TKE. 3D Smagorinsky, on the other hand, is coupled to a simpler surface exchange scheme based on \cite{Louis_1979_parametricmodelvertical}. Clearly, the difference in surface exchange schemes makes it difficult to compare the two configurations. At the same time, however, this comparison also helps us investigate the added value of the more complex surface exchange used with 1D TKE.

\section{Results}\label{sec3}
We are interested in days, when boundary-layer processes dominate and when a thermally-induced circulation is present in the Inn Valley. Thermally-induced flows were the subject of multiple observational and numerical studies \cite[e.g.,][]{ChowEtAl_2006_HighResolutionLarge,WeigelEtAl_2006_HighResolutionLarge,GogerEtAl_2018_ImpactThreeDimensional,KalverlaEtAl_2016_EvaluationWeatherResearch,LehnerEtAl_2019_MethodIdentifySynoptically,MikkolaEtAl_2023_Daytimevalleywinds} and can be considered as a well-known phenomenon of the mountain boundary layer. Thermally-induced flows with weak synoptic forcing are therefore an ideal test case for a process-based model validation.
\subsection{Process-based model evaluation}
We choose September 13, 2019 as a representative case study to give a detailed overview of ICON model performance. The general synoptic situation revealed a high-pressure ridge building up over central Europe, while a cut-off low was present over the Iberian Peninsula and a trough present over the Balkans. The synoptic flow over the Alps was therefore weak and mostly Northerly-Northwesterly. 
September 13, 2019 was one of the CROSSINN intensive observation periods (IOP), namely IOP8 \citep{AdlerEtAl_2021_CROSSINNFieldExperiment}, a cloud-free day with a well-developed up-valley flow in the afternoon. Asides from the LIDAR and eddy-covariance stations, radiosonde launches were performed every three hours starting from 03:00\,UTC, allowing us to conduct a detailed evaluation of the vertical structure of the valley atmosphere.
\subsubsection{Representation of the flow structure across grid spacings}
 \begin{figure}
 \noindent\includegraphics[width=\textwidth]{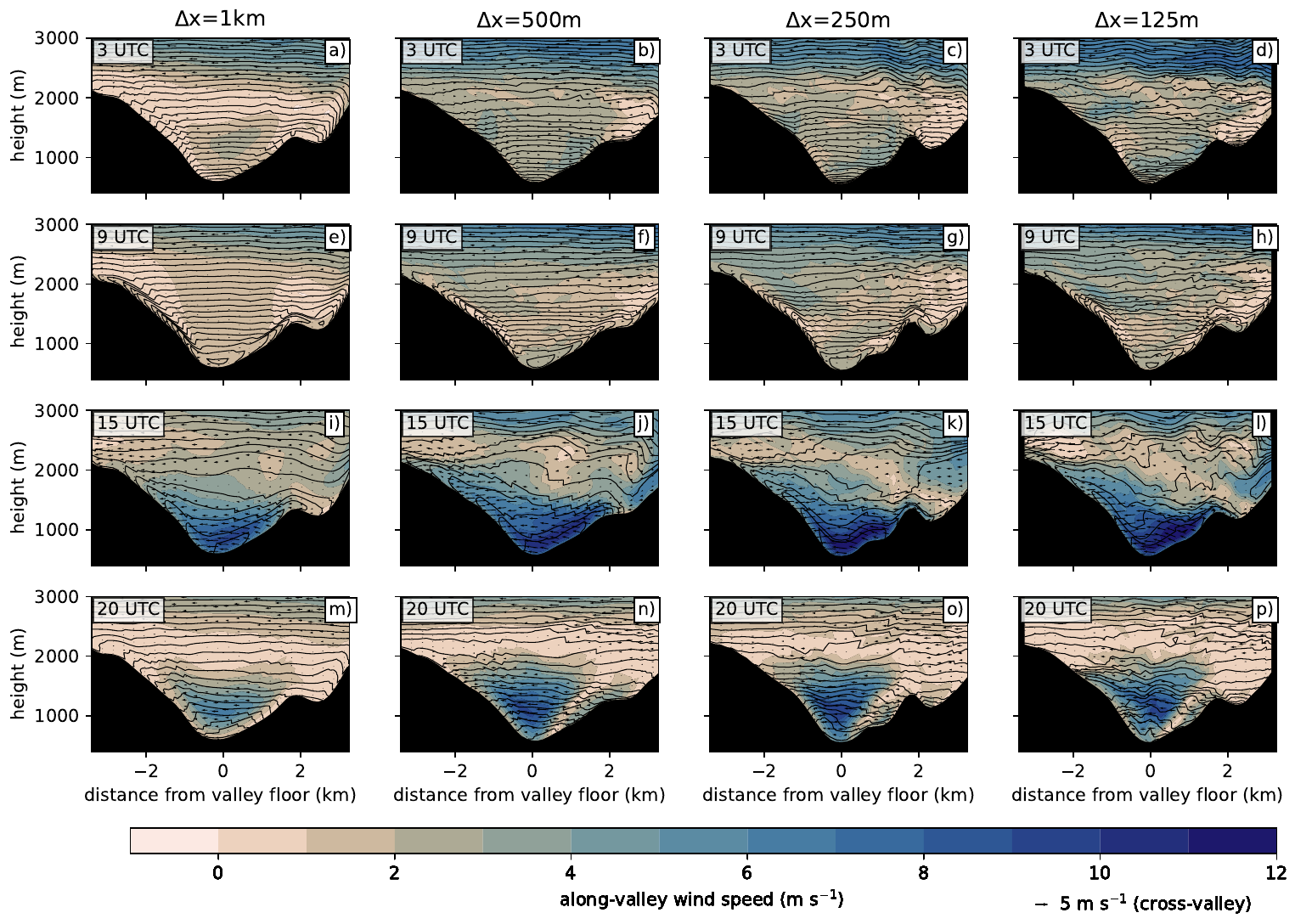}
\caption{Vertical cross-section following the purple line in Fig.~\ref{f:stations} from model output of potential temperature (contours) and along-valley flow speed (colors) with 3D Smagorinsky  across grid spacings (see Figure titles) from different times: 3\,UTC (panels a-d), 9\,UTC (panels e-h), 15\,UTC (panels i-l), 20\,UTC (panels m-p).}
\label{f:cs}
\end{figure}
At first, we investigate the representation of the thermally-induced circulation in the Inn Valley across the four grid spacings. 
In the night, at 03:00\,UTC (Fig.~\ref{f:cs}a-d), vertical cross-sections of simulated potential temperature reveal that a stable boundary layer (SBL) is present in the valley, with a cold-air pool (CAP) at the valley floor. Both along- and cross-valley wind speeds are generally low, with weak down-slope flows draining towards the valley floor. With decreasing $\Delta x$, the model is able to simulate a stronger CAP, especially visible in the lowest 500\,m above the valley floor in the DX-250 and DX-125 runs. The down-slope flows (Fig.~\ref{f:cs}a-d), are better resolved in the higher-resolution runs. The situation changes under the influence of incoming solar radiation: the nighttime SBL dissolved and a convective boundary layer (CBL) forms at the valley floor, visible in the unstable lowest layer at the valley floor with a vertical extent of around 500\,m above ground (Fig.~\ref{f:cs}e-h). The valley CBL is developed in most detail in the highest-resolution run, DX-125 (Fig.~\ref{f:cs}h). At both the North- and South-facing valley slopes, up-slope flows are visible across all resolutions, while the depth of the slope-wind layer varies: the deepest slope wind layer is visible at the DX-500 and DX-250 runs. In the afternoon, at 15:00\,UTC, the valley is mostly dominated by a strong up-valley flow (Fig.~\ref{f:cs}i-l) with a vertical extent of around 1000\,m above ground. The up-valley flow's intensity and spatial extent varies with horizontal grid spacing; in the DX-1000 run the along-valley wind speed maximum is around 9\,m\,s$^{-1}$, while in the higher-resolution runs, along-valley wind speed maxima of around 12\,m\,s$^{-1}$. Furthermore, a North-to-South cross-valley circulation is visible in all simulations and the up-slope flows were mostly eroded by the up-valley flow. In the evening (20:00\,UTC), the valley atmosphere stabilizes further and the up-valley flow weakens and lifts (Fig~\ref{f:cs}m-p). In the DX-1000 run, the valley floor is already mostly unaffected by the remains of the up-valley flow, while in the other simulations the up-valley flow is still stronger and still reaches the valley floor. 
\subsubsection{Vertical profiles of potential temperature and wind}
 \begin{figure}
 \noindent\includegraphics[height=\textheight, keepaspectratio]{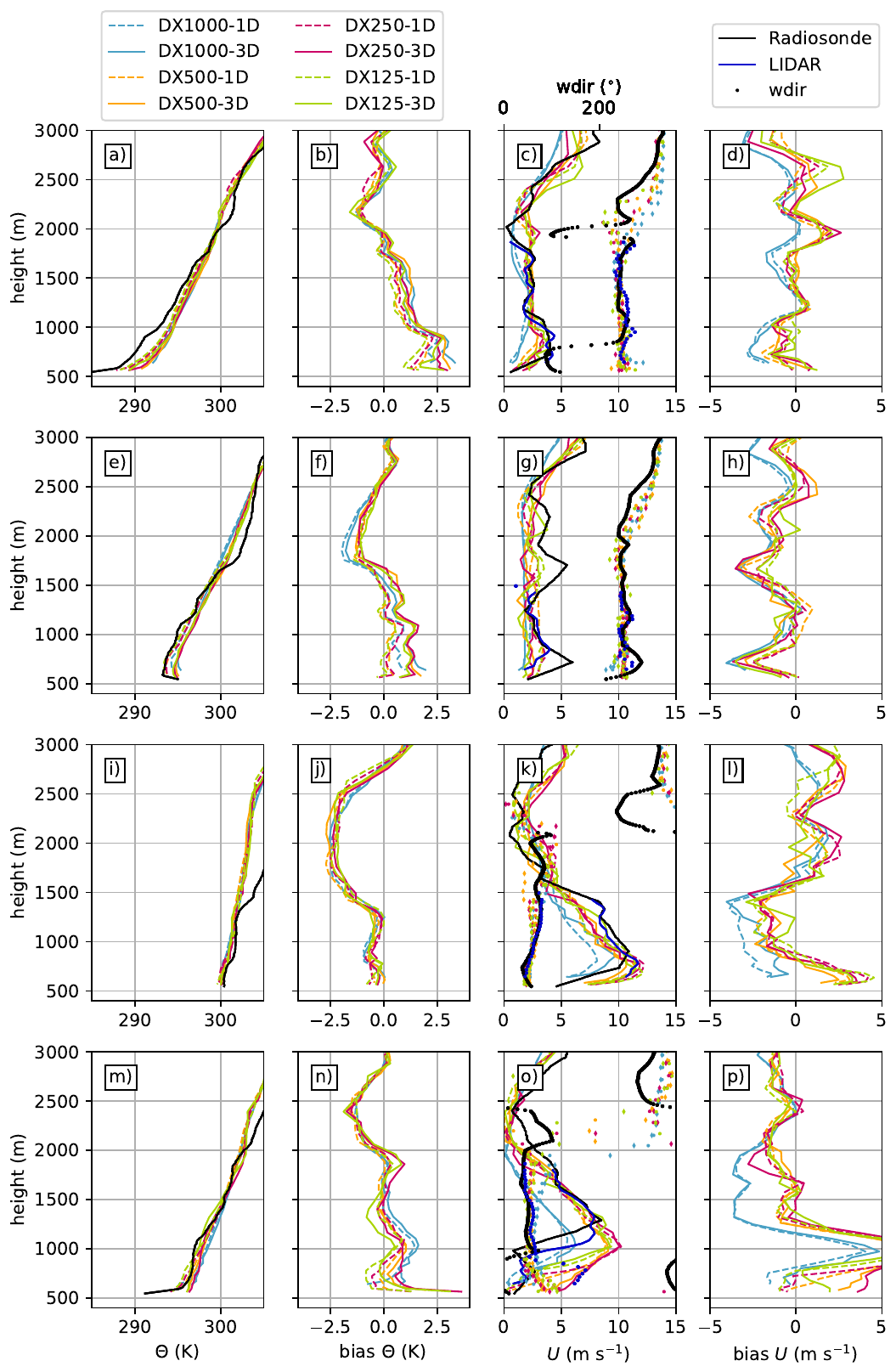}
\caption{Vertical profiles of potential temperature (first column; colors: model; black line: radiosonde observations), potential temperature bias (model-obs; second column), horizontal wind speed (third column; color lines: model; black line: radiosonde observations; dark blue line: LIDAR observations) and direction (third column; color dots: model; black dots: radiosonde observations), horizontal wind speed bias (model-obs; second column). Times: 03:00\,UTC (panels a-d), 09:00\,UTC (panels e-h), 15:00\,UTC (panels i-l), 20:00\,UTC (panels m-p). Dashed lines indicate simulations with the 1D TKE scheme, while full lines show simulations with the 3D Smagorinsky scheme.}
\label{f:raso}
\end{figure}
While the cross-sections of Figure~\ref{f:cs} give a useful overview of the flow structure simulated by the model, a more quantitative model validation is necessary. In the next section, we present a comparison of the computational results with radiosondes and LIDAR observations from the valley floor. \par
 The CAP and the nighttime SBL are also visible in the vertical profile of potential temperature observed by the radiosonde (Fig.~\ref{f:raso}a). The model simulates a CAP at the valley floor, but the potential temperature values are overestimated for all runs by around 2.5\,K in the lowest 1000\,m above ground (Fig.~\ref{f:raso}b). However, with decreasing $\Delta x$ (e.g., DX-125-1D), the model simulates a stronger CAP and the warm bias is reduced towards around 1\,K (Fig.~\ref{f:raso}b). This suggests that the model can simulate CAP formation more successfully at very high resolution (DX125) compared to the kilometric range (DX1000), even when the vertical grid spacing remains the same.  Still, the model is not able to capture the observed minimum potential temperature at the valley floor. Observed and simulated horizontal wind speeds are generally low during the night-time (Fig.~\ref{f:raso}c), while a low-level jet is visible in the lowest 550\,m of the radiosonde observations. The model underestimates this jet at the DX-1000 runs by 2\,m\,s$^{-1}$ (Fig.~\ref{f:raso}d), while the simulations at the hectometric range are able to represent the jet maximum both in magnitude and vertical extent. The radiosonde observations suggest an up-valley flow direction in disagreement with the LIDAR observations showing a down-valley flow. Given that in the night-time, down valley flows can be expected at this location, we assume that the up-valley flow direction from the radiosonde observations is likely a measurement artifact. The model simulates down-valley flows in agreement with the LIDAR observations (wind direction $\approx 270^{\circ}$). There are no large differences between the two turbulence schemes in the vertical atmospheric structure, but it has to be noted that the 3D Smagorinsky scheme exhibits a warm bias of around 1\,K compared to the 1D TKE scheme. \par
During the CBL phase, the radiosonde observations reveal a multi-layer inversion structure in the valley, where at least three distinct inversion layers are visible (Fig.~\ref{f:raso}e). This structure is related to subsidence and re-circulation zones of the up-slope flows, also observed in the Rivera Valley \citep{WeigelRotach_2004_Flowstructureturbulence}. The model simulates a mixed layer close to the ground as in the observations. Still, potential temperature is generally overestimated by the model, and the 3D Smagorinsky runs reveal a warmer mixed layer by about 1\,K than the 1D TKE runs (Fig.~\ref{f:raso}f). The simulation of the potential temperature profiles improves with decreasing grid spacing, and despite the warm bias, the potential temperature profile of DX125-3D shows a clearly visible multiple-layer valley atmosphere (Fig.~\ref{f:raso}e). The observations show a distinct second inversion around 1700\,m with warmer potential temperature aloft, which is not reproduced by the model. The observed and simulated wind speeds remain low (Fig.~\ref{f:raso}g), and the dominating wind direction is still down-valley. However, there is a slight disagreement below the mixed-layer height: the observations suggest a low-level wind speed maximum (with a turn in wind direction), which is missed by the simulations (Fig.~\ref{f:raso}g,h). \par
During the up-valley flow phase, the thermal structure of the valley atmosphere stabilizes and the vertical extent of the CBL is reduced (Fig.~\ref{f:raso}i). This stabilization is also simulated by the model and compared previous times (03:00 and 09:00\,UTC), the potential temperature bias is smallest (less than 1\,K, Fig.~\ref{f:raso}j). The distinct inversion lowered to 1500\,m, but is still not represented in the model. Concerning the structure of the up-valley flow (Fig.~\ref{f:raso}k), the up-valley maximum is weakest at DX1000-1D with an underestimation of the magnitude and the vertical extent of the jet (Fig.~\ref{f:raso}l). It should be noted that there is less underestimation with the DX1000-3D, likely because 3D Smagorinsky is a full three-dimensional closure and therefore also includes horizontal shear production beneficial for the realistic representation of the up-valley flow \citep{GogerEtAl_2018_ImpactThreeDimensional,GogerEtAl_2019_NewHorizontalLength}. The wind direction below crest height is Easterly, while above crest height (around 1000\,m above ground), the wind direction turns to Northerly, in accordance with the larger-scale synoptic flow and the plain-to-mountain circulation.  With decreasing $\Delta x$, the horizontal wind speed is not underestimated anymore; at DX500, the up-valley flow jet is almost perfectly simulated, while at DX250, it is even overestimated (Fig.~\ref{f:raso}l). \par
In the evening (20:00\,UTC), the up-valley flow weakens and lifts, and the valley atmosphere stabilizes further (Fig.~\ref{f:raso}m). This stabilization is also present in the simulated potential temperature profiles, but again with a warm bias, especially pronounced with more than 2.5\,K in DX250-3D and DX125-3D (Fig.~\ref{f:raso}n). The warmer valley atmosphere in the 3D Smagorinsky scheme likely prevents the growth of a shallow stable layer during the evening transition. However, the DX125-3D run manages to represent the adiabatic potential temperature layer below 1000\,m, which is not captured by DX125-1D. The correct simulation of the vertical extent of the up-valley flow jet during the evening transition is a challenge for the model (Fig.~\ref{f:raso}o): There is a large disagreement in horizontal wind speed of up to 5\,m\,s$^{-1}$ at 100\,m above ground (Fig.~\ref{f:raso}p). Furthermore, at the valley floor, only the runs with the 1D TKE scheme manage to simulate low wind speeds as the observations suggest indicating a delayed evening transition in the simulations using the 3D Smagorisnky scheme.
\subsubsection{Diurnal cycle of wind patterns}
 \begin{figure}
 \noindent\includegraphics[width=\textwidth]{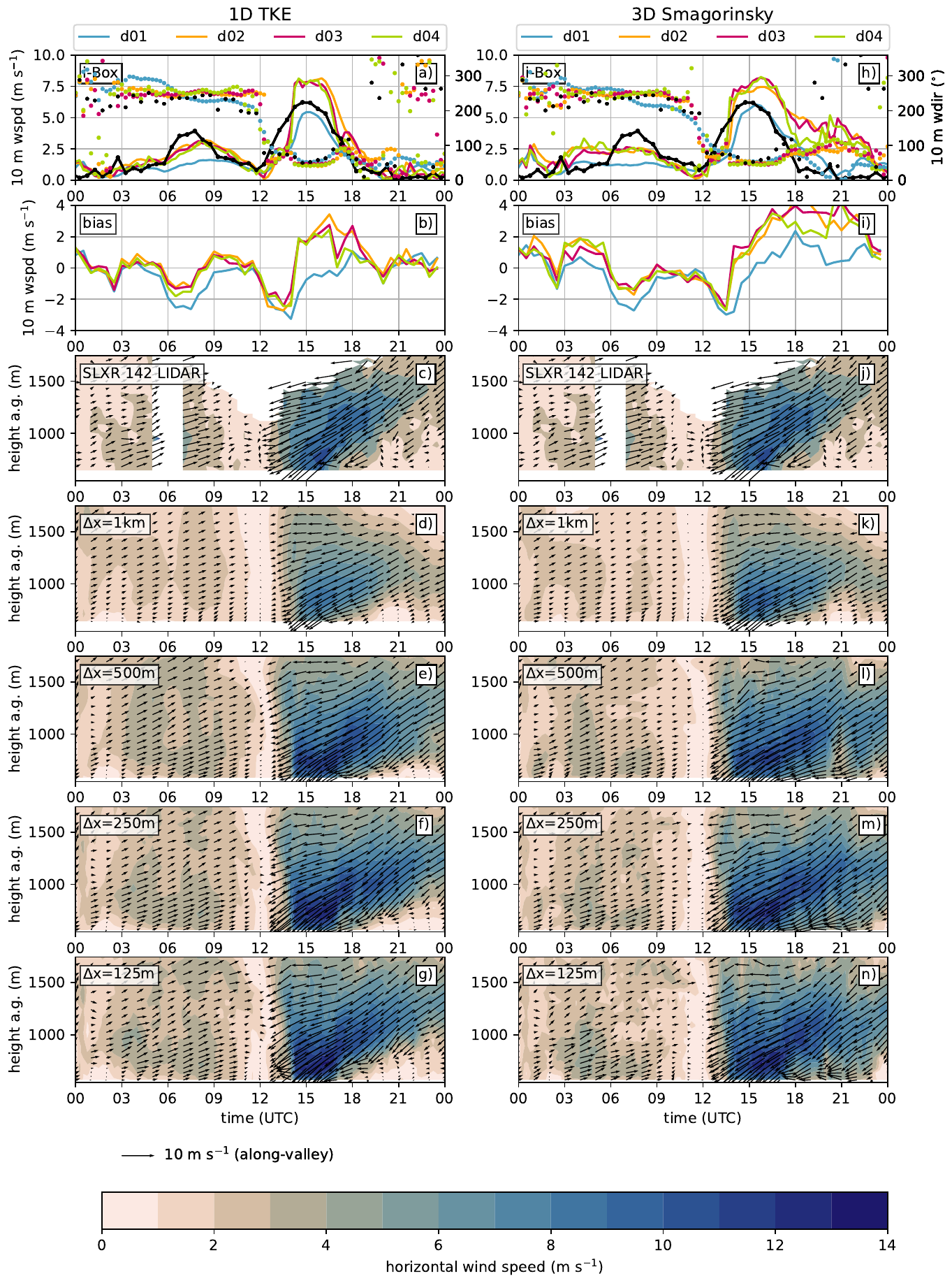}
 \caption{Panels a and h: Time series of wind speed and direction from observations from the i-Box flux tower located at the valley floor CS-VF0 (black lines/dots) with model output from the closest grid point (colors) for all domains from the 1D TKE scheme (panel a) and the 3D Smagorinsky scheme (panel h). Panels b and h: The bias (model-obs) of horizontal wind speed. Panels c and j: Observations of horizontal wind speed (colors) and wind direction (arrows) from the SLXR 142 Lidar located at the valley floor site. Corresponding time series from model output (colors) from the 1D TKE scheme (panels d-g) and the 3D Smagorinsky scheme (panels j-n).}
\label{f:lidar}
\end{figure}
The vertical cross-sections with the radiosondes only reveal snapshots of the valley atmosphere, therefore we analyze timeseries of simulated horizontal wind speeds together with vertical LIDAR profiles (Fig.~\ref{f:lidar}). Time series of observed wind speed and direction reveal that the valley floor is dominated by down-valley flows during the night-time (Fig.~\ref{f:lidar}a,b), successfully simulated by the model across resolutions and turbulence schemes. After sunrise, an observed (secondary) wind speed maximum is present at around 08:00\,UTC, while the wind direction remains down-valley. While DX1000 fails to simulate this phenomenon, the higher-resolution runs are able to capture the higher wind speeds before noon, although the timing of the maximum is shifted (Fig.~\ref{f:lidar}b,i). The onset of the up-valley flow occurs at 12:00\,UTC according to the observations, visible in the shift of wind direction from Westerlies (down-valley) towards Easterlies (up-valley). This up-valley flow onset is delayed by an hour in the simulations with the 1D TKE scheme; while the onset is almost on time in the 3D Smagorinsky scheme. Judging from the time series reveals an overestimation of 10\,m wind speeds for all hectometric grid spacings for both turbulence schemes (Fig.~\ref{f:lidar}b,i). An exception is DX1000-1D in which the 10\,m wind speed is simulated with a bias of less than 1\,m\,s$^{-1}$ during the up-valley flow phase. \par 
However, this view changes when we compare the vertical structure of DX1000-1D (Fig~\ref{f:lidar}d) with the LIDAR observations (Fig~\ref{f:lidar}c): The strength and vertical extent of the up-valley flow are underestimated by around 2\,m\,s$^{-1}$ in the DX-1000 simulations (Fig~\ref{f:lidar}d,k). The vertical spatial structure of the up-valley flow is closer to observations in the hectometric range (DX-500--DX-125, Fig~\ref{f:lidar}e-g,l-n). After 18:00\,UTC, the LIDAR observations show that the up-valley flow weakens and lifts (Fig~\ref{f:lidar}c,j), and the wind direction corresponds to up-valley flows except for the lowest levels, where there is already a shift towards down-valley flows (Fig~\ref{f:lidar}a,h). In the model, the schemes simulate different wind patterns: The 1D TKE scheme simulates the drop of horizontal wind speeds after 18:00\,UTC (Fig~\ref{f:lidar}e-g) in accordance with the observations, but the 3D Smagorinsky scheme simulates too high wind speeds close to the ground between 18:00\,UTC and 22:00\,UTC, suggesting that the evening transition is delayed (Fig~\ref{f:lidar}l-n). Therefore, we identify the decay of the up-valley flow in the evening in this case study as a challenge for the model - we will discuss the driving processes in the follow-up Section.

\subsubsection{Sensible heat fluxes and vertical velocities}\label{sec:et}

\begin{figure}
 \noindent\includegraphics[width=\textwidth]{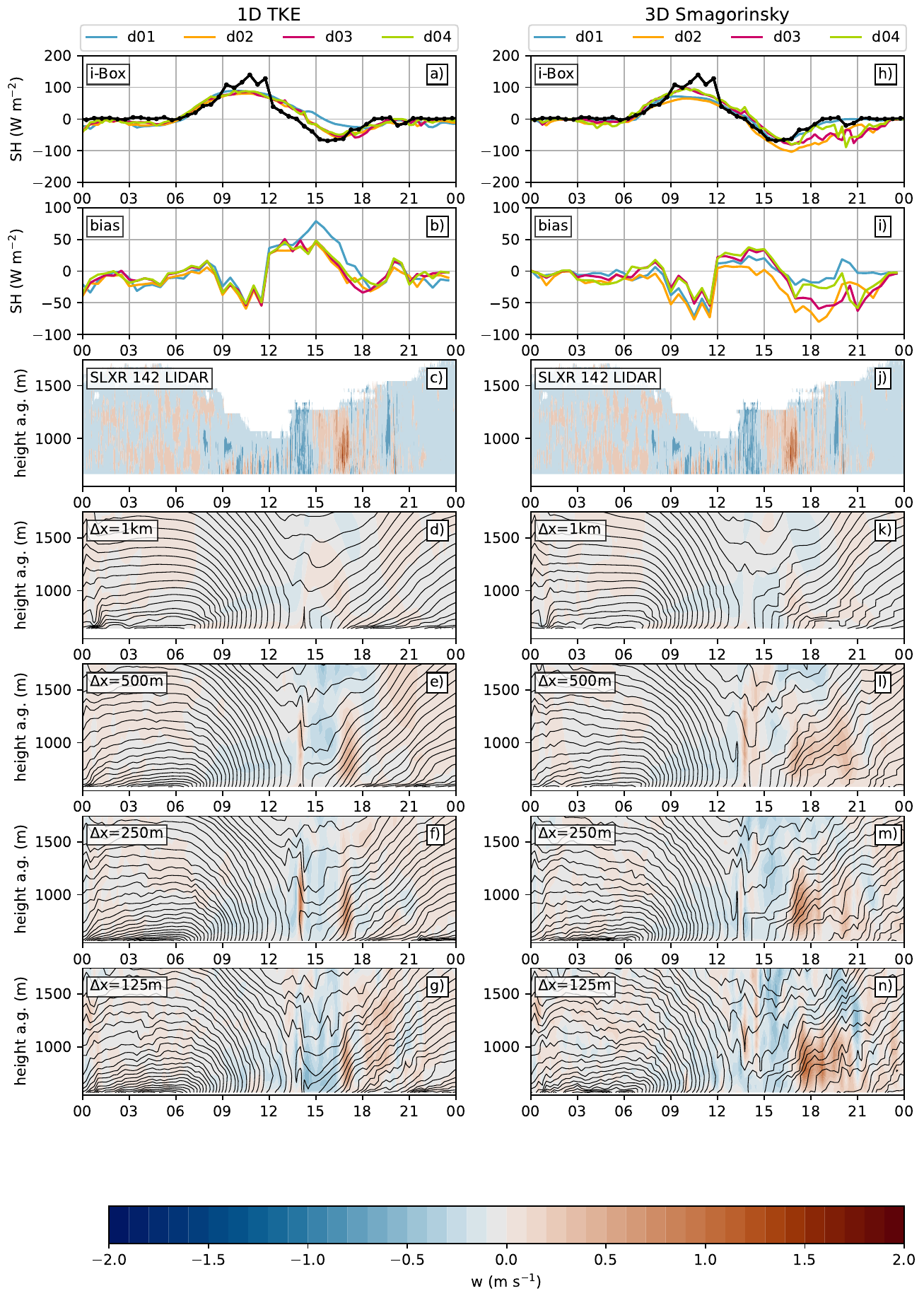}
 \caption{Panels a and g: Time series of sensible heat flux from observations from the i-Box flux tower at the valley floor CS-VF0 (black lines) with model output from the closest grid point (colors) from the 1D TKE scheme (panel a) and the 3D Smagorinsky scheme (panel g). Panels b and h: Observations  of vertical wind speeds (colors) from the SL 88 Lidar located at the valley floor. Panels c-f, i-l: Corresponding vertical velocity (colors) and potential temperature (black contours) time series from model output from the 1D TKE scheme (c-f) and the 3D Smagorinsky scheme (i-l).}
\label{f:lidar_w}
\end{figure}

The weakening of the up-valley flow and the transition back to down-valley flows poses a challenge for the model, and in the next paragraphs, we explore the reasons for this delay. One of the major forcings for ABL development is the sensible heat (SH) flux, and it is spatially heterogeneous in complex terrain \citep{MauderEtAl_2020_SurfaceEnergyBalance,LehnerEtAl_2021_Spatialtemporalvariations}. Time series from SH flux observations accompanied by model output across grid spacings (Fig.~\ref{f:lidar_w}a,h) reveal that SH fluxes are close to zero during the night-time (00:00-06:00\,UTC). After sunrise, the SH fluxes become positive and they already reach their daytime maximum before noon, when the CBL is well-developed (Fig~\ref{f:raso}e-h). During the CBL phase (09:00-12:00\,UTC), the simulations with the 3D Smagorinsky scheme (Fig.~\ref{f:lidar_w}i) exhibit larger SH fluxes (around 25\,W\,m$^{-2}$) than the 1D TKE counterparts (Fig.~\ref{f:lidar_w}b). The SH flux decreases again after 12:00\,UTC and turns negative at 14:00\,UTC, synchronous with the up-valley flow maximum (Fig.~\ref{f:lidar}), and this behaviour is a common feature in the Inn Valley and other Alpine valleys as well \citep{VergeinerDreiseitl_1987_Valleywindsslope,RotachEtAl_2008_Boundarylayercharacteristics,LehnerEtAl_2021_Spatialtemporalvariations,BabicEtAl_2021_Crossvalleyvortices}. The magnitude, the timing, and the change to negative values in the afternoon of the SH flux are represented successfully in both turbulence schemes. The timing of SH becoming negative at 14:00\,UTC is better represented in the 3D Smagorinsky scheme (Fig.~\ref{f:lidar_w}i), and there is a visible improvement in simulation with increasing resolution - while no such effect is visible in the 1D TKE scheme (Fig.~\ref{f:lidar_w}b). The largest discrepancy between observed and simulated SH fluxes in the 3D Smagorinsky scheme is visible during the evening transition (between 17:00-23:00 UTC in Fig.~\ref{f:lidar_w}h) in accordance with the overestimated horizontal wind speeds (Fig.~\ref{f:lidar}h,i), where negative SH fluxes are simulated, while the observations (and the 1D TKE scheme) suggest SH fluxes close to zero (Fig.~\ref{f:lidar_w}h,i).\newline
We now evaluate the vertical velocity behaviour at the valley floor to see if it corresponds to the SH structure. The SL 88 LIDAR at the valley floor performed scans in high temporal resolution of the wind components in vertical stare mode, including vertical velocities (Fig.~\ref{f:lidar_w}c,j). According to the observations, vertical velocities are small in the night-time in the first morning hours until around 08:00\,UTC. After 08:00\,UTC, the CBL at the valley floor develops with several smaller up- and downdrafts until around 13:00\,UTC. When the up-valley flow arrives, the vertical velocities are mostly negative associated with subsidence at the valley floor, while after 15:00\,UTC, mostly positive vertical velocities dominate. A distinct vertical velocity maximum is visible at around 16:00\,UTC, interestingly, when the SH flux reaches its minimum and the evening transition of the up-valley flow starts. During the evening transition, vertical velocities are mostly dominated by up- and downdrafts with a larger vertical extent than during the CBL phase before noon. After 21:00\,UTC, vertical velocities become very small again.\par
The DX1000 runs only simulate a very crude vertical velocity structure for both turbulence schemes (Fig.~\ref{f:lidar_w}d,k). This is not surprising, since it can not be expected that small-scale up- and downdrafts are resolved at $\Delta x$=1\,km. DX500-1D and DX250-1D show a more detailed vertical velocity structure (Fig.~\ref{f:lidar_w}e-f); suggesting subsidence during the CBL phase and an updraft when the up-valley flow arrives. This strong, unrealistic updraft might contribute to the overall too large horizontal and vertical extent of the simulated up-valley flow at these grid spacings (Fig.~\ref{f:lidar}e-f). At the start of the evening transition, DX500-1D and DX250-1D simulate the strong updraft at around 16:00\,UTC correctly. The DX500-3D and DX250-3D runs (Fig.~\ref{f:lidar_w}i-m) show a similar picture as their 1D counterparts, however, after the initial evening transition updraft, we note continuous positive vertical velocity values after 17:00\,UTC, while in the observations, there are subsequent up- and downdrafts present. The evening transition in the 3D Smagorinsky runs is dominated by positive vertical velocities and the isentropes show that the growth of a SBL is prevented (Fig.~\ref{f:lidar_w}j-n). \par

\begin{figure}
 \noindent\includegraphics[width=\textwidth]{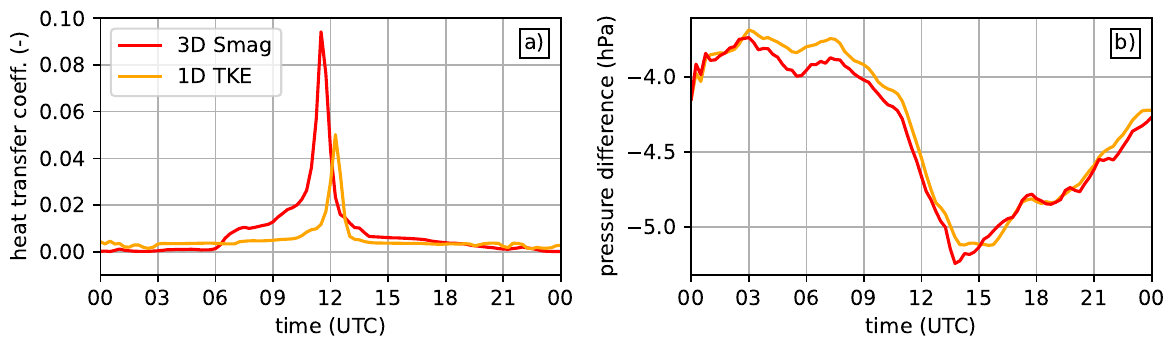}
 \caption{Time series of (a) the dimensionless surface exchange coefficient for heat at the valley floor (CS-VF0) and (b) of the pressure difference at the lowest model level between the valley floor and the Inn Valley entrance. Only DX500 is shown, but the behavior is similar for all other resolutions.}
\label{f:exchange_coeff}
\end{figure}

 Finally, we know from the vertical profiles (Fig.~\ref{f:raso}) that the potential temperature at the lowest levels is around 1\,K higher in the 3D Smagorinsky scheme during the CBL phase in the valley (Fig.~\ref{f:raso}f). This ``warmer CBL'' is related to higher SH fluxes between 9\,UTC and 12\,UTC than in the 1D TKE scheme (Fig.~\ref{f:lidar_w}b,i). The higher SH flux results from larger heat transfer coefficients between the surface and the atmosphere (Fig.~\ref{f:exchange_coeff}a), which is due to the difference in the way surface exchange is handled in the two schemes. The warmer valley atmosphere leads to a stronger pressure gradient between the valley and the surrounding Alpine foreland (Fig.~\ref{f:exchange_coeff}b). This allows a build-up of a stronger valley-wind circulation, especially visible in the vertical profiles of horizontal wind speed during the afternoon and evening (Figs.~\ref{f:raso}o,p). Therefore, the most likely reason for the delayed evening transition in the 3D Smagorinsky scheme happens several hours before the evening transition: A warmer valley boundary layer leading to a stronger pressure gradient results in a stronger up-valley flow, which prevails longer than in the simulations with the 1D TKE scheme (delayed evening transition, Fig.~\ref{f:lidar}g-l). 
 
\subsubsection{Spatial variability of wind patterns}

\begin{figure}
 \noindent\includegraphics[width=\textwidth]{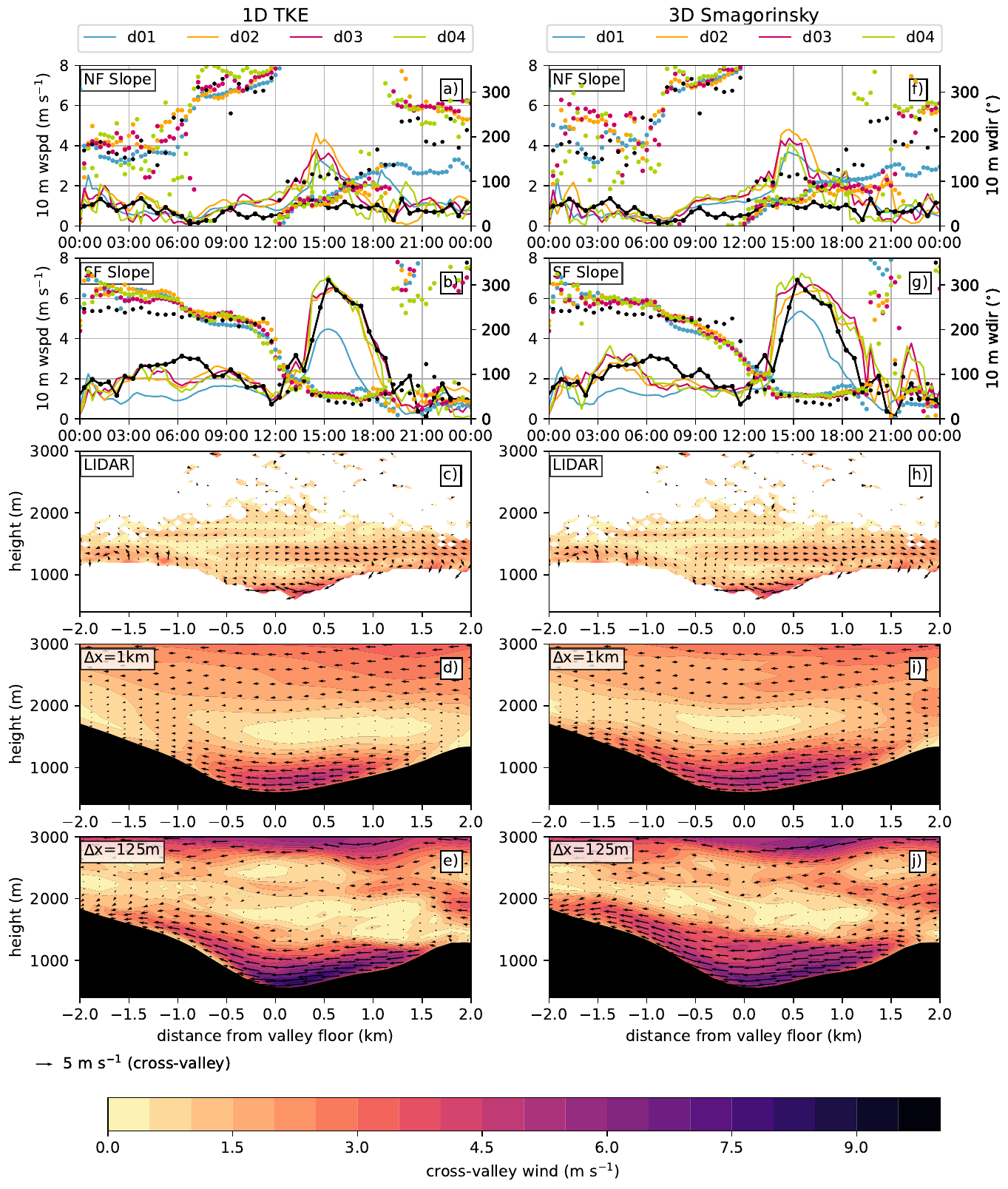}
 \caption{Panels a,b and f,g: Time series of wind speed and direction from observations from the i-Box flux towers located at the North-facing slope CS-NF27 (black lines/dots, panels a,b) and the South-facing slope CS-SF1 (black lines/dots, panels b,g) with model output from the closest grid point (colors) for all domains from the 1D TKE scheme (panels a,b) and the 3D Smagorinsky scheme (panels f,g). Panels c and h: Coplanar scans at 15:00\,UTC from the Lidars located at the North-facing slope (CS-NF27), the valley floor (CS-VF0), and the South-facing slope (Mairbach), with observed cross-valley flow speeds (colors and wind arrows). Corresponding model output from DX-1000 and DX-125 and the two turbulence schemes (Panels d-e: 1D TKE; panels i-j: 3D Smagorinsky).}
\label{f:cop}
\end{figure}
In the previous Section we showed the model performance at the valley floor. However, in complex terrain, it is likely that one single point measurement is not representative for the boundary-layer structure in the entire valley. Therefore, we introduce time series of observations and model output across resolutions from the North-facing slope (CS-NF27, Fig~\ref{f:cop}a,f) and the South-facing slope (CS-SF1, Fig~\ref{f:cop}b,h). \par
The wind speed observations at the North-facing slope station reveal a completely different structure than at the valley floor station (Fig.~\ref{f:lidar}a,b): Horizontal wind speeds remain low (smaller than 2\,m\,s$^{-1}$) during the entire night and day. The wind direction switches from down-slope before 06:00\,UTC (ca. 180$^{\circ}$) to up-slope flows until 12:00\,UTC (ca. 340$^{\circ}$). After midday, the wind directions shifts towards up-valley (ca. 50-90$^{\circ}$) while the wind speeds remain low, indicating that the up-slope flows are eroded by the larger-scale up-valley flow. The erosion of slope flows was also observed in the Rivera Valley \citep{RotachEtAl_2008_Boundarylayercharacteristics} and is a common phenomenon at this particular slope station \citep{GogerEtAl_2018_ImpactThreeDimensional}. The correct simulation of the slope flow erosion depends on $\Delta x$: At DX1000, the up-valley flow is simulated too weakly, as shown in the analysis of the (vertical) structure in the previous Section. However, the model simulates a wind maximum in the afternoon, synchronous with the up-valley flow maximum at the valley floor (Fig~\ref{f:cop}a,f). The afternoon maximum is more pronounced in the two intermediate domains (DX500 and DX250). These results suggest that the model simulates a too strong slope flow erosion by the up-valley flow - and therefore struggles to simulate the scale interaction between the small-scale slope flows and the larger-scale up-valley flow correctly.  \par
The South-facing slope station (Fig~\ref{f:cop}b,g) shows similar wind patterns as at the valley floor. Down-valley flows dominate during the night-time and before noon, and the up-valley flow arrives at 12:00\,UTC with a sharp increase in wind speed and a shift in wind direction towards Easterlies. The model underestimates the up-valley flow at DX1000-1D by 2\,m\,s$^{-1}$, while DX1000-3D is more successful in the simulation of the wind speed maximum (Fig~\ref{f:cop}g). In general, the wind speed maximum is represented at the hectometric range, while the 3D Smagorinsky scheme manages to simulate a sharper wind speed increase with the arrival of the up-valley flow. As at the valley floor station, the evening transition with the drop of the wind speed after 18:00\,UTC is delayed in the runs 3D Smagorinsky scheme, although the wind direction shifts back to down-valley earlier than at the valley floor.\par
To understand the representation of the cross-valley wind patterns in the model better, we show cross-sections of simulated cross-valley wind speed together with wind observations from LIDAR co-planar scans (Fig.~\ref{f:cop}c-e,h-j). LIDAR observations at 15:00\,UTC (valley flow maximum) reveal a wind speed maximum (4\,m\,s$^{-1}$) at the valley floor with North to South cross-valley flow (Fig.~\ref{f:cop}c,h). Between the heights of 500\,m and 1500\,m above ground, a cross-valley vortex with a re-circulation pattern is visible. The cross-valley vortex of the Inn Valley is the result of both thermal and dynamical forcing and forms due to a balance by the pressure gradient force and the centrifugal force due to the valley's curvature \citep{BabicEtAl_2021_Crossvalleyvortices}. The model is not able to reproduce this detailed cross-valley flow pattern at DX1000 (Fig.~\ref{f:cop}d,i): Although the modelled and observed cross-valley flow speeds and directions match, the model simulated a larger vertical extent of the wind speed maximum up to 1000\,m above ground, while the observations reveal the re-circulation zone at this height. Furthermore, the horizontal extent of the cross-valley flow is larger - which, finally, leads to the too strong up-valley flow influence on the North-facing slope station. 
At the highest resolution (DX125, Fig.~\ref{f:cop}e,j), the cross-valley flow is represented in more detail (esp. in the 3D Smagorinsky scheme, Fig.~\ref{f:cop}j) with a very weak re-circulation pattern visible around 2000\,m.

\subsection{Model validation over five case studies}
\subsubsection{2\,m temperature and 10\,m wind speed (i-Box stations)}
We introduced a detailed, process-based model validation in the previous sections from a single case study. However, there is still the question whether the conclusions from one case study can be applied to the other five cases as well. The three i-Box station along the valley cross-section (Fig.~\ref{f:stations}) give an overview whether the model can account for spatial variability and if changing the horizontal grid spacing yields better results for certain locations. Therefore, we calculated the root-mean square error (RMSE) for two common meteorological variables (2\,m temperature and 10\,m horizontal wind speed) as in \cite{ChowEtAl_2006_HighResolutionLarge} and \cite{GogerEtAl_2018_ImpactThreeDimensional},
  \begin{equation}
   rmse=\left(\frac{1}{N_t}\sum^{N_t}_{i=1}{(M_{i}-O_i)}^2\right)^\frac{1}{2},
  \end{equation}
  where $M_i$ is the median of the five case study days, $O_i$ are the i-Box observations, and $N_t=5$ are the number of case study days.\par
\begin{figure}
  \noindent\includegraphics[width=\textwidth]{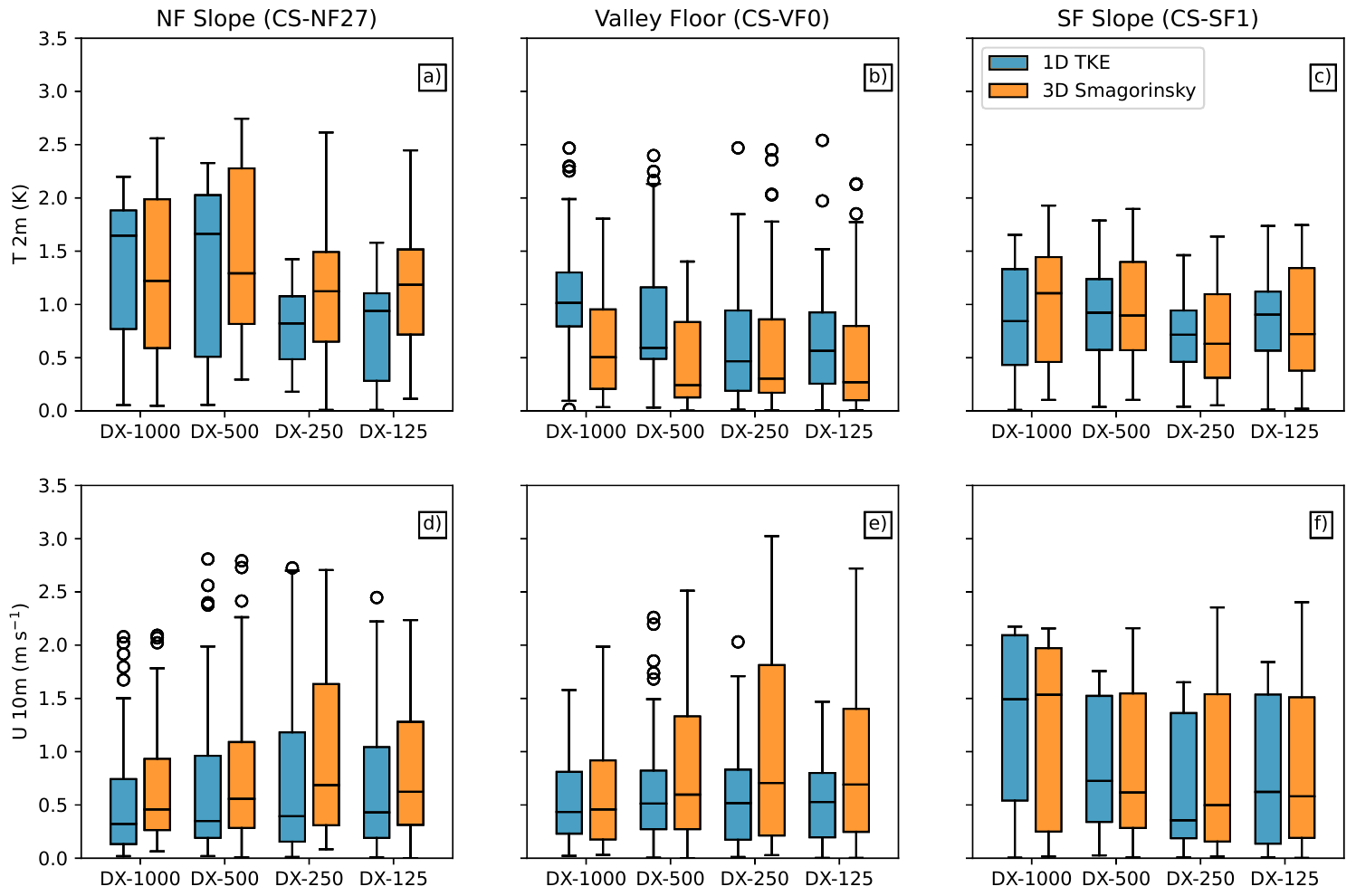}
  \caption{RMSE calculation of the average over all five case study days at the North-facing Slope station (CS-NF27, left column), the valley floor station (CS-VF0, middle column), and the South-facing Slope station (CS-SF1, right column) of  2\,m temperature (a-c) and 10\,m wind speed (d-f). Blue boxes indicate simulations with the 1D TKE scheme, while orange boxes represent the 3D Smagorinsky scheme at the respective horizontal resolutions. The straight lines in the box plot the median, the box represents the 25\,th and 75\,th percentiles, the whiskers represent the 10\,th and 90\,th percentiles, and the circles represent outliers.}
 \label{f:veri}
 \end{figure}

 The results across resolutions from the three i-Box locations (NF slope, valley floor, and SF slope) are shown in Figure~\ref{f:veri}.
The RMSE for 2\,m temperature (Fig.~\ref{f:veri}a-c) suggests that the largest RMSE are present at the NF slope station (around 0.5\,K larger than the other two stations). This is not surprising, given the challenging location and the difficulties of the model to represent scale interactions accordingly (cf. Fig.~\ref{f:cop}). Increasing the horizontal resolution improves the simulation of 2\,m temperature for all three investigated locations, but the NF slope station benefits most. In general, the reduction of the RMSE by grid spacing are more striking for the 1D TKE scheme (e.g., a reduction of RMSE of -0.5\,K at the NF slope station from DX-1000 to DX-125, Fig.~\ref{f:veri}a). However, the 2\,m temperature RMSE is smaller for the 3D Smagorinsky scheme across resolutions and stations (e.g., Valley floor DX-125-3D shows -0.25\,K compared to the 1D TKE run, Fig.~\ref{f:veri}b). If we would only judge from the RMSE of 2\,m temperature, the 3D Smagorinsky scheme clearly outperforms the 1D TKE scheme for the valley floor and the SF slope, while at the NF slope station, increasing the resolution is the major factor to improve the 2\,m temperature simulation. \par
A different picture emerges from the 10\,m wind speed RMSE (Fig.~\ref{f:veri}d-f): Horizontal resolution has a smaller impact on the reduction of RMSE at all stations, the only exception is the SF slope station where RMSE in the  DX-1000 runs is more than 1.5\,m\,s$^{-1}$ larger than the hectometric range counterparts. At all locations, the 3D Smagorinsky scheme simulates higher wind speeds than the 1D TKE scheme, at the valley floor, this is mostly related to the delayed evening transition discussed previously. In case of the 10\,m wind speed no definite conclusions can be drawn whether horizontal grid spacing or the turbulence scheme contribute to an improved model simulation. 

\subsubsection{Diurnal cycle of horizontal wind speed (LIDAR observations)}
  \begin{figure}
  \noindent\includegraphics[width=\textwidth]{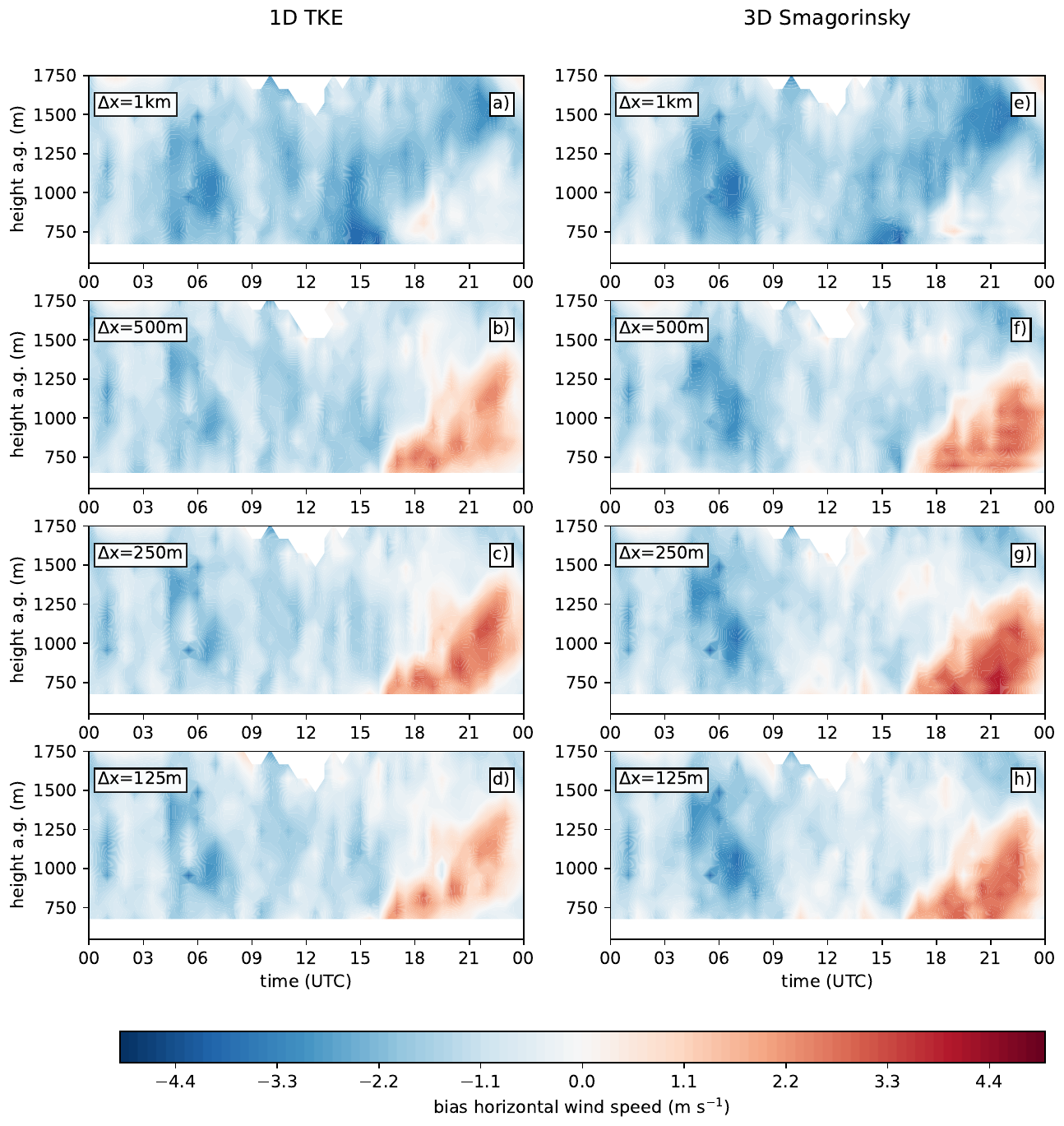}
  \caption{Bias calculation from vertical profile time series of the average over all five case study days of simulated horizontal wind speed versus LIDAR observations  at the valley floor.}
 \label{f:veri_lidar}
 \end{figure}
With the aid of the LIDAR observations in vertical stare mode (cf. Fig.~\ref{f:lidar}), we can also calculate the averaged model bias of the diurnal cycle of horizontal wind speed over five case study days (Fig~\ref{f:veri_lidar}). The averaged LIDAR data were interpolated to the model grid so that the calculation of the two-dimensional model bias is possible.\par
The averaged wind speed bias suggests that horizontal wind speed is underestimated by the model for the "average" valley wind day, across resolutions and turbulence schemes. The DX-1000 simulations exhibit the largest negative bias of -4\,m\,s$^{-1}$ during the up-valley flow phase at around 15\,UTC (cf. Fig.~\ref{f:lidar}a), while the bias is smaller for the 3D Smagorinsky run (bias: -3\,m\,s$^{-1}$). The negative wind speed bias is reduced for all simulations in the hectometric range. After 17\,UTC, when the up-valley flow is breaking down and the evening transition starts, the model has a positive bias of up to +4\,m\,s$^{-1}$. While, as expected, the largest positive bias values are present for the simulations with the 3D Smagorinsky scheme (e.g., DX-500-3D, Fig.~\ref{f:lidar}g), it is interesting to note that the hectometric simulations with the 1D TKE scheme also show a positive bias during the evening transition as well. This finding was not evident from the analysis of the single case study, but apparently, the decay of the up-valley flow is a general challenge for the model, independent of the turbulence and/or surface transfer scheme.

\section{Discussion}\label{sec4}
The present study shows a detailed overview of the performance of the ICON model at the hectometric range for the thermally-induced circulation. Generally speaking, the model delivers a realistic simulation of the flow structure in the valley across all tested grid spacings. The largest jump in model improvement is visible from the kilometric (DX1000) towards the hectometric (DX500) range: While time series of the wind speed suggest no large differences, the hectometric range simulations especially improve the vertical structure of the up-valley flow jet (Figs.~\ref{f:lidar},\ref{f:veri_lidar}). Furthermore, a more realistic representation of the thermal structure of the valley atmosphere is visible at the hectometric range, especially during the CAP phase in the night-time. In accordance, the RMSE of 2\,m temperature benefits most from the increase in horizontal resolution (Fig.~\ref{f:veri}). This is in agreement with findings by \cite{SchmidliQuimbayoDuarte_2023_DiurnalValleyWinds} and \cite{TianEtAl_2024_stationbasedevaluation}, who noted a generally improved model performance in the hectometric range compared to similar kilometric simulations.\par
However, there are still physical processes posing a challenge for the model: The most striking problem in the model is the delayed evening transition in the simulations. A detailed analysis of several boundary-layer variables suggests that the too high sensible heat flux in the 3D Smagorinsky scheme before noon leads to a too strong up-valley flow, which also decays slower than in the 1D TKE counterpart (with a different surface transfer scheme). \cite{BantaEtAl_2023_MeasurementsModelImprovement} found similar results in the HRRR model, where a too high sensible heat flux leads to a too strongly simulated sea-to-mountain breeze in the Western US. Transitional boundary-layer regimes are often a challenge for NWP models \citep{CouvreuxEtAl_2016_Boundarylayerturbulent} and even large-eddy simulations \citep{StollEtAl_2020_LargeEddySimulation}. However, we identified the major issue in ICON and in the future, a careful assessment of the surface transfer scheme will be necessary; furthermore, the correct representation of surface fluxes has a larger impact on the correct boundary-layer simulation than the turbulence scheme.\par
Accounting for surface heteorgeneity is still one of the major challenges for NWP models \citep{CalafEtAl_2023_BoundaryLayerProcesses}. The model is generally able to account for surface heterogeneity due to its high horizontal resolution. However, the simulated heterogeneity of the flow field is still limited, mostly visible at the North-facing slope station: The model simulated a too strong up-valley flow, leading to a too early erosion of the local slope flow. Scale interactions in complex terrain are a common phenomenon, but models still struggle with their correct representation, mostly because usually the ``smaller-scale'' process is not simulated correctly \citep{UmekEtAl_2021_Largeeddysimulation,AdlerEtAl_2023_Evaluationcloudycold}. \par
Comparing the two turbulence schemes with each other resulted in a generally similar model performance. Differences were detectable in the simulation of the up-valley flow, where the 3D Smagorinsky scheme generally improved the onset of the timing the up-valley flow and its vertical structure. This could be related to the three-dimensionality of the 3D Smagorinsky scheme, because the up-valley flow in the Inn Valley is a phenomenon where horizontal shear already plays a role at a length scale of around 1\,km \citep{GogerEtAl_2018_ImpactThreeDimensional,GogerEtAl_2019_NewHorizontalLength}. Although the 3D Smagorinsky scheme simulates a ``too warm'' boundary layer before noon, the scheme is able to simulate a more detailed thermal structure in the valley, which also reflects in the 2\,m temperature statistics, where it outperforms the 1D TKE scheme. Therefore, we found no specific reason to completely dismiss the 3D Smagorinsky scheme, especially because the additional three-dimensional effects in the scheme might lead to a more realistic representation of three-dimensional mountain boundary layer processes.\par
As a last point, we stress that most findings of this study would have been impossible without the high-resolution observations of the CROSSINN campaign. Our model analysis goes beyond a comparison of grid box results with point observations, where only assumptions, but no actual conclusions could be drawn from. 
In our results, we would not have found that the valley wind jet is too weak in the kilometric simulations, the improved representation of vertical velocities in the DX-125 runs, or the too strong cross-valley circulation in the model leading to misrepresented scale interactions in the model (to name a few). Multi-dimensional observations are therefore essential to validate models at the hectometric range, and a process-based model validation would not be possible with single point observations from operational observational networks. Measurement campaigns are, however, limited in time, but the year-long TEAMx campaign \citep{RotachEtAl_2022_CollaborativeEffortBetter} will take place in the near future, providing a major opportunity to continue detailed model validation over mountainous terrain.

\section{Conclusions and Outlook}\label{sec5}
In this study, we investigated the performance of the ICON model across four horizontal grid spacings ($\Delta x=1$\,km, $\Delta x=500$\,m, $\Delta x=250$\,m, and $\Delta x=125$\,m) for the simulation of the mountain boundary layer in the Inn Valley, Austria. Furthermore, we employed two available turbulence schemes - a 1D boundary-layer parameterization (1D TKE) and a fully 3D turbulence closure (3D Smagorinsky) to assess the performance of the two schemes across resolutions. For model validation, we utilized radiosonde profiles, LIDAR scans (vertical profiles and cross-valley scans), and point observations from the i-Box turbulence flux towers. We simulated five case studies of days in summer/autumn of 2019, where a thermally-induced circulation with up-valley flows is present, and we analyzed one case study from CROSSINN IOP8 in detail. The results allow us to draw the following conclusions:
\begin{enumerate}
\item Generally speaking, the model is successful in simulating the valley boundary layer structure across horizontal grid spacings and turbulence schemes. 
\item The largest improvement in model performance is visible in the jump from $\Delta x=1$\,km to $\Delta x=500$\,m: At the hectometric range, the model shows a more realistic vertical structure of the up-valley flow and an improved simulation of 2\,m temperature.
\item The process-based model validation with the CROSSINN observations allowed us to identify the challenges for the model for the simulation of thermally-induced flows, namely a delayed evening transition and scale interactions.
\item Differences between the two turbulence schemes emerge due to the different surface transfer schemes: Too high sensible heat fluxes in the 3D Smagorinsky scheme lead to a too strong up-valley flow and henceforth a delayed evening transition.
\item The 3D Smagorinsky scheme at $\Delta x$=125\,m is able to simulate a more realsitic and detailed structure of the valley boundary layer than the 1D TKE scheme, especially visible in the vertical profiles and vertical velocities.
\item The model is able to represent the spatial variability of the valley boundary layer, although it tends to simulate a too strong cross-valley circulation, leading to the erosion of small-scale slope flows at the slopes.
\item General model validation statistics (RMSE) for all five case studies from three stations suggest that a switch towards the hectometric range is beneficial for the simulation of the thermally-induced circulation, and the improvement is more pronounced for the 2\,m temperature than for the 10\,m wind speed.
\end{enumerate}
This study aimed to give an overview whether simply increasing the horizontal resolution yields better model results. The findings agree that the lower boundary condition - namely, the better-resolved topography - mainly governs the boundary-layer flow. The CROSSINN observations provide a valuable data pool for a detailed model evaluation for the valley cross-section, but our model evaluation is very localized and still limited to the lower Inn Valley. We will extend the analysis of our simulation dataset in the near future by exploring the turbulence representation in the model across scales and by identifying the issues in the surface transfer scheme coupled to the 3D Smagorinsky closure.

\section*{acknowledgements}
This research was supported by the EXCLAIM project funded by the ETH Zurich. The computational results presented have been achieved using resources from the Swiss National Supercomputing Centre (CSCS) under project ID d121.

\section*{conflict of interest}
The authors declare no conflict of interest.

 \section*{data availability}

 Observational data stem from the CROSSINN campaign \citep{AdlerEtAl_2021_CROSSINNFieldExperiment}. The SL88 and SLXR 142 LIDAR data were retrieved from \cite{gohm_alexander_2021_4585577}. The coplanar LIDAR scans and radiosonde data were downloaded from \cite{AdlerEtAl_2021_CROSSINNCrossvalley,AdlerEtAl_2021_CROSSINNCrossvalley2}. The post-processed turbulence flux tower data from the i-Box stations can be retrieved at \url{https://acinn-data.uibk.ac.at/}.  We used the ICON model code version 2.6.5 for our simulations. The open-source model code can be obtained at \url{https://icon-model.org/}. Figures were generated with python-matplotlib \citep{Hunter_2007_Matplotlib2DGraphics} using colormaps by \cite{crameri_fabio_2023_8035877}.

\printendnotes

\bibliography{qjrms_goger_bibliography}

\end{document}